\documentclass[final,twocolumn,prl,aps,floats,amsfonts,amssymb,showpacs,superscriptaddress]{revtex4}
\usepackage[utf8]{inputenc}
\usepackage{graphicx}
\usepackage{color}
\usepackage{epstopdf}
\usepackage{amssymb}
\usepackage{amsmath} 

\definecolor{orange}{rgb}{0.9,0.55,0.3}
\definecolor{navyblue}{rgb}{0,0,0.8}
\definecolor{forest}{rgb}{0,0.8,0}
\definecolor{lightblue}{rgb}{0.6,0.8,1}
\definecolor{garnet}{rgb}{0.8,0,0}
\definecolor{teal}{rgb}{0.04,0.52,0.78} 
\definecolor{grey07}{rgb}{0.7,0.7,0.7}
\definecolor{turq}{rgb}{0,0.75,0.75}

\newcommand{\st}[1]{$(\widetilde{#1})$}
\newcommand{\alert}[1]{\textcolor{red}{#1}}

\begin{document}

\title{Forcing-type-dependent stability of steady states in a turbulent swirling flow}

\author{B. Saint-Michel} 
\email{brice.saint-michel@cea.fr}
\author{B. Dubrulle}
\affiliation{Laboratoire SPHYNX, Service de Physique de l'\'Etat Condens\'e, DSM, CEA Saclay, CNRS URA 2464, 91191 Gif-sur-Yvette, France}
\author{L. Marié}
\affiliation{Laboratoire de Physique des Océans, UMR 6523 CNRS/IFREMER/IRD/UBO, Brest, France}
\author{F. Ravelet}
\affiliation{Laboratoire Dynfluid, ENSAM ParisTech, CNRS EA92, 151, boulevard de l'H\^{o}pital 75013 Paris, France}
\author{F. Daviaud$^1$}


\begin{abstract}

We study the influence of the forcing on the steady turbulent states of a von K\'arm\'an swirling flow, at constant impeller speed, or at constant torque. We find that the different forcing conditions change the nature of the stability of the steady states and reveal dynamical regimes that bear similarities with low-dimensional systems. We suggest that this forcing dependence may be an out-of-equilibrium analogue of the \emph{ensemble inequivalence} observed in long-range interacting statistical systems, and that it may be applicable to other turbulent systems.

\end{abstract}

\maketitle

\paragraph*{Introduction}
An intriguing property of statistical systems with long range interactions is the \emph{ensemble inequivalence}: a solution in the microcanonical (constrained) ensemble is not necessarily a solution in the canonical (unconstrained) ensemble~\cite{Thirring1970,Ellis2000}. This property traces back to the non-additivity of energy, and is reflected by pathological behaviours of the entropy (that can be non-concave) or the heat capacity (that can become negative). Ensemble inequivalence has been observed and studied in a variety of systems such as for example 2D Euler equations~\cite{Venaille2009,Herbert2012}, Blume-Emery-Griffiths model~\cite{InequivEnsemble},  random graphs~\cite{Barre2007}. More recently, it has also been studied in a model describing the one-dimensional motion of N rotators coupled through a mean-field interaction, and subjected to the perturbation of an external magnetic field~\cite{Ninno2012}; it shows that this concept may also hold out of equilibrium, broadening its range of applicability. Nevertheless, a generalisation of such concepts to all out-of-equilibrium systems remains ---  at present time --- an open question. We propose an extension of such tools to turbulent systems. They naturally display long-range interactions and are, by definition, far from equilibrium. Our experimental system is a ``von K\'arm\'an" experiment, in which a cylinder of fluid is stirred by two impellers fitted with 8 curved blades, producing fully-developed turbulence in a relatively small experimental device. Several features of equilibrium systems have already been described in this model experiment, such as steady states, predicted by the resolution of the axisymmetric Euler equations~\cite{Monchaux2006}, hysteresis~\cite{Ravelet2004}, and spontaneous symmetry-breaking with diverging susceptibility~\cite{Cortet}.


In this letter, we will examine the stability of such steady states under two different forcing conditions, either imposing the \emph{speed} or the \emph{torque} ---~flux of kinetic momentum~\cite{Marie2004} --- to our impellers. The subject has attracted little attention, work focusing on the difference of power fluctuations~\cite{Titon2003,Leprovost2004,Bramwell1998} under both conditions. However, we can consider our forcings to be conjugate, as the product torque $\times$ speed controls the energy injection rate in our experiment. Switching from speed to torque control might then be seen as an analogue of switching from \emph{canonical} to \emph{microcanonical} ensemble; it is shown to alter the stability of the steady states previously observed in~\cite{Ravelet2004} and to reveal interesting dynamical regimes.

\paragraph*{Experimental setup}

The von K\'arm\'an flow is created in a polycarbonate cylinder of inner radius $R = 100$~mm filled with water. The fluid is stirred by two 8-bladed impellers of radius $0.925R$ separated from each other by a distance (blade-tip to blade-tip) $1.4R$. Two independent $1.8$~kW brush-less motors can rotate the impellers either by imposing their speeds $(f_1,f_2)$ or their torques $(C_1,C_2)$. In addition to the AC drive measures, torque and speed measurements are performed by two Scaime MR12 torque sensors fixed to the mechanical shafts driving the impellers. Fluid confinement is assured by two balanced mechanical seals under a $2.8$~bar pressure to provide minimum friction. Temperature control is regulated by an external water flow in two refrigeration coils installed behind each impeller. The Reynolds number of our experiment, defined as $Re = \pi (f_1 + f_2) R^2 / \nu$, and varying from $2\cdot 10^5$ to $5\cdot 10^5$, is well above the transition to turbulence reported in~\cite{Ravelet2008}.

\begin{figure*}
	\includegraphics[width = \textwidth]{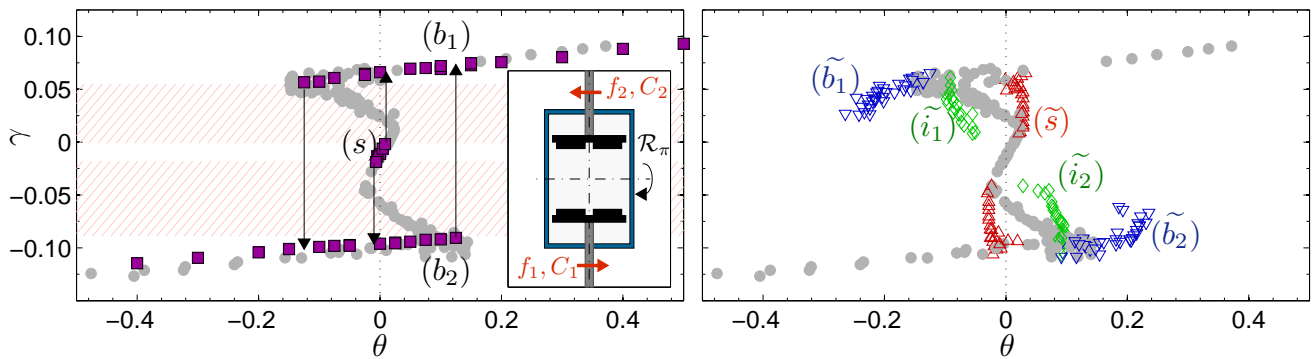}
	\caption{(Left), mean reduced torque asymmetry $\gamma$ plotted as a function of the mean reduced speed asymmetry $\theta$, for both speed (purple squares) and torque (grey circles) experiments. The arrows indicate the possible transitions between steady states, sketching a hysteresis cycle which include a forbidden $\gamma$ zone (hatched region) for \emph{speed control} experiments. No hysteresis is observed in~\emph{torque control}. (Right), modes of the $\theta$ probability density function for torque control experiments corresponding to the `forbidden range'. \st{s}, \st{b_1} and \st{b_2} are branches of quasi-steady states extending respectively the steady $(s)$, $(b_1)$ and $(b_2)$ branches. \st{i_1} and \st{i_2} are new branches of quasi-steady states never observed in speed control. (Inset) Sketch of the `VK2' experiment, with the two impellers (black). The experiment is axisymmetric along the vertical axis, and is $\mathcal{R}_\pi$ symmetric for exact counter-rotation.}
	\label{fig:gammatheta}
\end{figure*}

Our experiments aim to measure the response of the flow to asymmetric forcing for both types of controls. \emph{Speed} control experiments will impose $(f_1 + f_2)/2 = 4$~Hz, to study the influence of $f_1 \neq f_2$ on the values of $C_1$ and $C_2$. Reciprocally, \emph{torque} control fixes $(C_1 + C_2)/2 = 1.72$~Nm to study the effect of torque asymmetries $C_1 \neq C_2$ on the impeller speeds $f_1$ and $f_2$. The upside-down $\mathcal{R}_\pi$ symmetry (see inset in fig~\ref{fig:gammatheta}) provides a definition of two antisymmetric dimensionless quantities $\theta$ and $\gamma$ to measure the forcing --- and response --- asymmetry: $\theta = (f_1 - f_2)/(f_1 + f_2)$ is the reduced impeller \emph{speed} difference and $\gamma = (C_1 - C_2)/(C_1 + C_2)$ is the reduced shaft \emph{torque} difference.

\paragraph*{Speed control} 

For speed-imposed experiments, all turbulent flows are steady. These steady states are characterized by their mean torque asymmetry $\gamma$. Starting both impellers at the same time for $\theta$ very close to 0, the system reaches steady states corresponding to a `symmetric' branch called $(s)$. They statistically consist of two Ekman recirculation cells separated by a shear layer, in agreement with~\cite{Ravelet2004,Cortet}. In such states, a small $\theta$ variation triggers a transition with a dramatic increase of the mean torque. These new `bifurcated' states exhibit only one circulation cell, and form two separate branches of the $(\gamma, \theta)$ plane (see the left part of fig.~\ref{fig:gammatheta}). These branches, named $(b_1)$ and $(b_2)$, respectively exhibit a global pumping of the bottom or the top impeller. Velocimetry measurements have confirmed that the velocity fields of the flows belonging to the $(b_1)$ and $(b_2)$ are images of each other by the $\mathcal{R}_\pi$ rotation, restoring globally the experimental symmetries. Once on $(b_1)$ and $(b_2)$ branches, it is never possible to reach the $(s)$ state, which is therefore marginally stable. In addition, the $(b_1)$ and $(b_2)$ branches are hysteretic, $(b_1)$ states persisting for $f_1 \leq f_2$ and $(b_2)$ states for $f_2 \leq f_1$~\cite{Remark}, agreeing with previous results~\cite{Ravelet2004}. This hysteresis is associated with a `forbidden zone' of $\gamma$ values never accessed for imposed speed.

\begin{figure}
	\includegraphics[width = 0.48\textwidth]{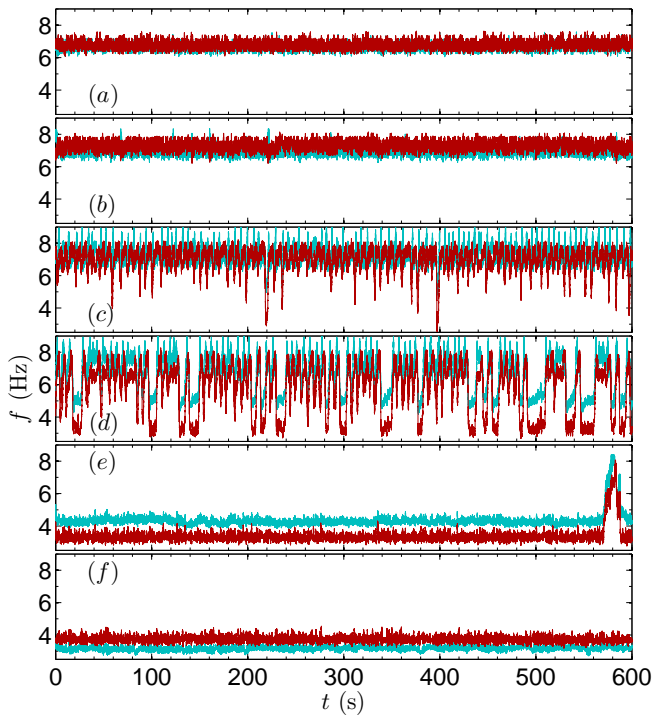}
	\caption{Temporal series of the impeller speeds $f_1$ (\textcolor{turq}{blue}) and $f_2$ (\textcolor{garnet}{red}) for various  $\gamma$. $(a)$, symmetric high-speed state \st{s} observed at $\gamma = -0.0164$; $(b)$, threshold of the irregular oscillations \st{i_2} with very small events for $\gamma= -0.0460$; $(c)$, \st{i_2} irregular oscillations for $\gamma = -0.0668$; $(d)$, multi stable regime showing \st{s}, \st{i_2} and \st{b_2} events at $\gamma = -0.0891$; $(e)$, a single fast rare event in a quasi-steady slow \st{b_2} regime at $\gamma = -0.0912$; $(f)$, slow $(b_2)$ regime for $\gamma = -0.1049$.}
	\label{fig:tempor}
\end{figure}

\paragraph*{Torque control}
In contrast, imposing \emph{torque} allows \emph{any} value of $\gamma$ assuming friction is negligible. We have first verified that imposing $\gamma$ out of the `forbidden zone' provides steady states identical to those observed in speed control, following the three branches already described in fig.~\ref{fig:gammatheta}. Such flows are almost identical to speed-control flows, no difference in velocimetry measurements being visible after suitable normalization. Our experiments have then focused on the henceforth accessible `forbidden zone'. In this region, the system looses its steadiness: the speed of the impellers may alternatively jump between multiple attractor turbulent states. This multi stability is identified by the emergence of multiple peaks in the P.D.F. (probability density function) of the $1.5$~Hz low-pass filtered signal of $\theta(t)$. Such filtering is required considering our speed measurements generate discrete values of the impeller speed; it yields a robust density function when the filter cut-off frequency is changed. Three types of attractor states have then been defined: \st{s}, the high-speed state, with characteristics similar to $(s)$, \st{b_1} and \st{b_2} which are low-speed states similar to $(b_1)$ and $(b_2)$, and two new \st{i_1} and \st{i_2} \emph{intermediate} states. These new states can be seen in fig.~\ref{fig:gammatheta}; while \st{s}, \st{b_1} and \st{b_2} states extend well beyond their speed-imposed counterparts, the \st{i_1} and \st{i_2} branches are new and cannot be observed in speed control. Decreasing $\gamma$ from a perfectly symmetric \st{s} ($\theta = 0$) state, we can observe its influence on the temporal signals, (cf. fig.~\ref{fig:tempor}). First, steady states with decreasing mean $\theta$ are observed. Then (fig.~\ref{fig:tempor}$(b)$), when $\gamma \leq -0.049$, corresponding to a local extremum of the mean value of $\theta$, small localized peaks of $f_1$ and $f_2$ are simultaneously observed, breaking time invariance. Such events are identified as excursions towards intermediate state \st{i_2}. Still decreasing $\gamma$, the peaks grow until the biggest events saturate at low $f_1$ and $f_2$ (fig.~\ref{fig:tempor}$(c)$). These events are identified as transitions to the \st{b_2} state. For even lower values of $\gamma$, the system behaviour is irregular, constantly switching between fast \st{s}, \st{i_2} and slow \st{b_2} states (fig.~\ref{fig:tempor}$(d)$). In this situation, all states are found to be quasi-steady, each of them being able to last more than $10$~sec. ($70$~impeller rotations). Decreasing further $\gamma$ affects the dynamics of the system, more time being spent in \st{b_2} at the expense of \st{i_2} and \st{s}. Therefore, for low $\gamma \geq -0.0920$ (fig.~\ref{fig:tempor}$(e)$), only rare events can drive the system to the faster \st{i_2} and \st{s} states. Eventually, for $\gamma \leq -0.099$ (fig.~\ref{fig:tempor}$(f)$), time-invariance of the system is restored, corresponding to a $(b_2)$ steady state of fig.~\ref{fig:gammatheta}. Remarkably, the flow susceptibility defined by the mean value of $\theta$, $\partial \gamma / \partial \theta$, is \emph{negative} in all this forbidden zone (see fig.~\ref{fig:gammatheta}). Obviously, \emph{increasing} $\gamma$ progressively from a perfectly symmetric state leads to the same sequence of events, though \st{i_1} and \st{b_1} will be reached.

\begin{figure}
	\includegraphics[width = 0.48\textwidth]{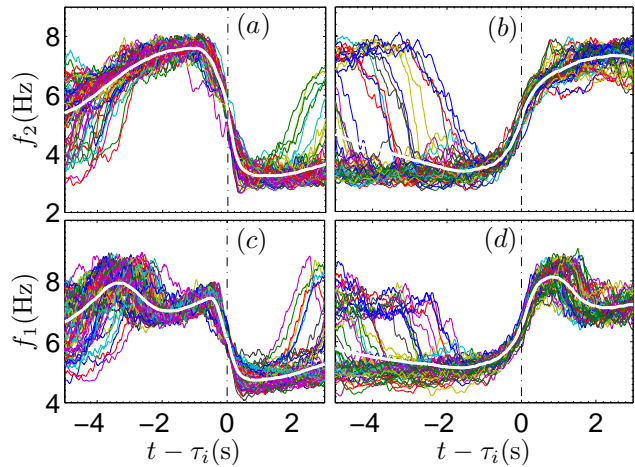}
	\caption{Shifted temporal signals of $60$ randomly-chosen transitions of a 2-hours experiment with $\gamma = -0.0891$. We compute $\tau_i$ by finding the minimum of $|\partial_t \bar{f_{2}}|$ where $\bar{.}$ means the $.$ signals numerical filtering at $1$~Hz. $(a)$, $(c)$, respectively $f_2$ and $f_1$ profiles for down transitions. $(b)$, $(d)$, respectively $f_2$ and $f_1$ for up transitions. The thick white line represents in each sub-plot the rotation frequency averaged on all $195$ events of the experiment.}
	\label{fig:mytransitions}
\end{figure}

\begin{figure}
	\includegraphics[width = 0.48\textwidth]{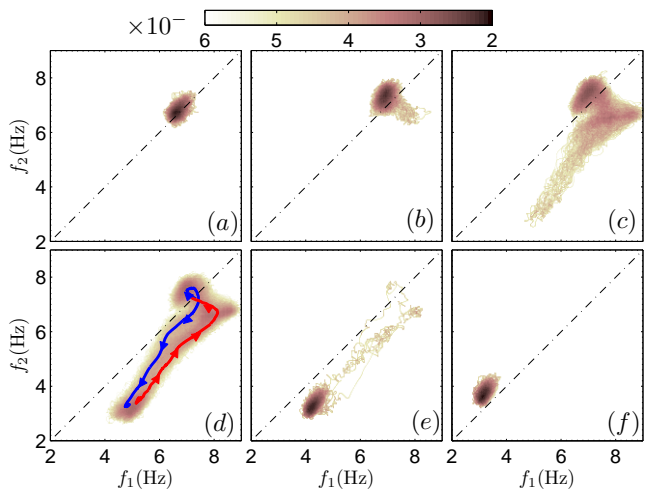}
	\caption{Joint-probability density maps of the $(f_1,f_2)$ values (log-scale), based on the temporal signals of fig.~\ref{fig:tempor}: $(a)$, steady symmetric high-speed state: $\gamma = -0.0164$; $(b)$, beginning of irregularly oscillating states towards an intermediate state, $\gamma = -0.0460$; $(c)$, typical irregular oscillations with large slowdowns: $\gamma = -0.0668$; $(d)$, multistability with three `most visited' states: $\gamma = -0.0891$, (\textcolor{blue}{---}) and  (\textcolor{red}{---}) represent respectively the fig.~\ref{fig:mytransitions} mean profile for down and up transitions; $(e)$, rare events for $\gamma = -0.0912$; $(f)$, steady slow state: $\gamma = -0.1042$. The black dashed-dotted line represents the $\theta = 0$ condition. Both fast and slow states can be observed at $\theta = 0$ considering the shape of the fig.~\ref{fig:gammatheta} curve.}
	\label{fig:hists2d}
\end{figure}

Valuable information on our system dynamics can be found studying the temporal signals of the global quantities near the transitions~\cite{Berhanu2007}. We have therefore superposed in fig.~\ref{fig:mytransitions} the speed signals close to the transitions observed in fig.~\ref{fig:tempor}$(c)$: \st{s} $\to$ \st{b_{1,2}} is called a down transition, and \st{b_{1,2}} $\to$ \st{s} an up transition. Once the transition instant accurately determined, a good collapse of all curves is observed, validating a unique transition path. This extends the low-dimensional system description of~\cite{Berhanu2007} to purely hydrodynamical quantities in a non-magnetic turbulent flow.
Eventually, the joint distributions of $(f_1, f_2)$ are studied to highlight the attractor states that emerge from figs.~\ref{fig:gammatheta} and~\ref{fig:tempor}. In fig.~\ref{fig:hists2d}$a$), for $\gamma$ too small, only one peak appears, which confirms the steady nature of the system in $(s)$. For small asymmetries, small excursions that escape the attractor, corresponding to the previously described small \st{i_2} oscillations can be found, exhibiting a new peak strongly deviating from the diagonal $f_1 = f_2$. For higher asymmetries, the system fills a large part of the $(f_1, f_2)$ plane, with three main maxima: \st{s} close to the diagonal at higher $(f_1,f_2)$, \st{b_2} for the off-diagonal low $(f_1, f_2)$ regime. The third --- \st{i_2} --- attractor is harder to see because it is partially hidden by neighbouring zones repeatedly crossed by unsteady events. It is located between the right tip of the histogram and the \st{s} maximum. With this representation, one clearly observes that the mean system path is different for the down and up transitions: while the down transition seems to start ``looping" in the vicinity of \st{s} before abruptly transiting to \st{b_2}, the up transition reaches the right edge of the joint-PDF $(f_1>f_2)$, near \st{i_2} before joining the \st{s} state. 

The repartition of the height of the maxima in fig.~\ref{fig:hists2d}$(c)$, $(d)$, $(e)$ is obviously driven by $\gamma$, from almost-fully \st{s}, \st{i_2} to nearly-pure \st{b_2} with rare large transitions to the faster states. For nearly-pure \st{s}, we clearly see (fig~.\ref{fig:hists2d}$(c)$) that a large amount of small excursions occur, contrasting with the nearly-pure \st{b_2} state (fig~.\ref{fig:hists2d}$(e)$), for which a small number of large events is reported. The position of such maxima is \emph{not} fixed when $\gamma$ varies: we recall for example that fig.~\ref{fig:gammatheta} shows a clear variation of the position of the steady $(s)$ and $(b_1)$ and $(b_2)$ maxima with respect to $\gamma$. Interestingly, while the \st{b} and \st{s} absciss\ae  increase with $\gamma$, the \st{i} position \emph{decreases} with $\gamma$.

\begin{figure}[h]
	\centering
	\includegraphics[width = 0.48\textwidth]{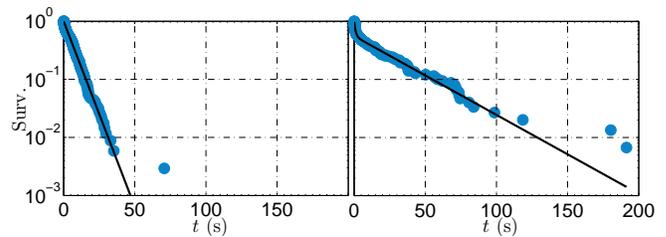}
	\caption{Survival function $1 - \int p.d.f.$ of the distribution of the time spent in the slow quasi-steady state \st{b_1} before transiting to one of the faster quasi-steady states \st{i_1} and \st{s}. (\textcolor{teal}{$\bullet$}), experimental data, broken line, exponential fit with two characteristic times. Left, $\gamma = -0.094$ favours a simple exponential distribution. Right, $\gamma = -0.097$, closer to the edge of the steady $(b_2)$ branch, favours separate characteristic times.}
	\label{fig:survival}
\end{figure}

\paragraph*{Discussion}
Using global torque and speed measurements, we have characterized the response of the von K\'arm\'an experiment to different energy injection mechanisms. The two responses coincide in the range of parameters accessible to speed control, reproducing the hysteresis cycle previously reported by~\cite{Ravelet2004}. However, imposing the torque $\gamma$ in the zone which could not be reached in speed control generate new continuous ``mean" branches connecting symmetric $(s)$ and bifurcated $(b)$ branches. The mean values of the speed asymmetry $\theta$ hide the underlying phenomena observed in this forbidden zone, revealing multiple peaks of the P.D.F of $\bar{\theta}$, each peak corresponding to a quasi-steady state. Two of them can be defined by continuity of the steady, speed-control branches: \st{s} and \st{b}. The third state, \st{i}, is never observed in any speed-imposed experiment. The study of the impellers velocity $f_1(t), f_2(t)$ signals shows typical excursions and transitions between our three states, similarly to~\cite{Petrelis2008, Petrelis2009}, while preliminary results on the distribution of \st{b} residence time favour exponential Kramers-like escape times~(see fig.~\ref{fig:survival}) where the longer characteristic time increases when approaching the bifurcated $(b_2)$ steady branch. This confirms a ``potential well" interpretation of the quasi-steady states as previously performed by~\cite{Burguete2007}. 

%
\alert{
Our results address several questions. The most striking result is the multi-valued asymmetry response curve, $\gamma(\theta)$. Interestingly, solid state devices can \emph{also} display similar multivalued characteristic curves $i(U)$~\cite{Esaki1958}. It is therefore tempting to draw an electrical analogy between the von K\'arm\'an flow and an electrical dipole, where the flux of angular momentum through the flow, fixed by setting the motor torques, would be analogous to the flow of electric charges through the dipole, the role of the applied voltage $U$ being played by the impeller speeds. In that respect, our system bears similarities with ``bulk" negative differential resistances~\cite{Ridley1963}. However, in all cases, negative resistances are associated with a spatial phase separation which cannot be sustained by our hydrodynamical, strongly correlated experiment.}

\alert{More generally, from the point of view of statistical physics, the von K\'arm\'an experiment allows a quantitative analysis of the influence of the energy injection on the response of an out-of-equilibrium system. In that respect, negative responses are characteristic of long-range interacting (LRI) systems, where ensemble inequivalence has been studied~\cite{Dauxois2003}. For such systems, switching from micro-canonical ensemble, with fixed energy density $e$, to canonical ensemble where the temperature $T$ is imposed, can result in different sets of equilibrium states. Micro-canonical solutions can interestingly display stable regions with a negative specific heat $c_v = \partial e / \partial T$ --- another analogue of negative susceptibility $\partial \gamma / \partial \theta$ and negative conductivity $\partial i / \partial U$ --- . In contrast, canonical solutions always impose $c_v \geq 0$ to avoid thermal reservoir instability~\cite{InequivEnsemble}. Imposing the impeller speed (or the applied voltage) can therefore be seen as a canonical exploration of the system response, whereas imposing the torque (or the current) is an equivalent of micro-canonical ensemble exploration, for which negative average susceptibilities (or resistances) are allowed. Eventually, the dynamic multi-stability observed in the forbidden zone can be seen as a probing of metastable quasi-steady states due to out-of-equilibrium turbulent noise, or as temporal inhomogeneities in a strongly correlated system for which no ``spatial phase separation" is accepted.}


From the point of view of turbulence, the multivalued region sets the problem of \emph{universality} of the steady states, that appears to be rather sensitive --- on large scales --- to the energy injection mechanisms, at variance with traditional view of turbulence. The phenomenon we explore in the present letter could actually be present in other turbulent experimental systems: for example, turbulent Plane-Couette flows could be forced either with constant global stress (motor torque $C$) or global strain (speed $f$), Poiseuille flows have been studied either imposing a pressure difference or a mass flow-rate~\cite{Darbyshire1995}, and even Rayleigh-Bénard experiments have been conducted under temperature-imposed and the heat-flux imposed conditions~\cite{Johnston2009}.
\alert{Previous work on rotating Rayleigh-Bénard convection, which can be viewed as analogous to our $\theta \neq 0$ experiments, have actually shown interesting ``chaotic" dynamics~\cite{RotConv}. We eventually need to interrogate the influence of the forcing nature in natural systems where reversals are observed: northern hemisphere winds show alternative zonal and blocked~\cite{BlockZon} patterns}, whereas Kuroshio currents~\cite{Kawabe1995} might deviate abruptly from their initial course: both systems exhibit complex dynamics with jumps between quasi-steady states.

\paragraph*{Acknowledgements}
We thank the CNRS and CEA for support, Vincent Padilla for building the experimental device, Guillaume Mancel for sharing experimental data, and Cécile Wiertel-Gasquet for writing the acquisition programs.


\begin{thebibliography}{14}
\bibitem{Remark} This range, as well as fig.~\ref{fig:gammatheta} do not exactly respect $\mathcal{R_\pi}$ symmetry. It is actually very difficult to obtain accurate symmetric calibrations because such experiments require small and accurate torques.
\end{thebibliography}
\end{document}